# A Multi Perspective Approach for Understanding the Determinants of Cloud Computing Adoption among Australian SMEs

*Research in Progress*

Salim Alismaili
[1]School of Computing and Information Technology
Faculty of Engineering and Information Systems
University of Wollongong
Szaai787@uowmail.edu.au

Mengxiang Li[1]
mli@uow.edu.au

Jun Shen[1]
jshen@uow.edu.au

Qiang He
School of Software and Electrical Engineering
Faculty of Science, Engineering & Technology
Swinburne University of Technology
qhe@swin.edu.au

## Abstract

Cloud computing is proved to be an effective computing technology for organisations through the advantages that it offers such as IT technical agility and scalability, enhancing businesses processes, and increasing enterprises competitiveness. In Australia, there is an emerging trend that small and medium-sized enterprises (SMEs) begin to adopt this technology in the conventional working practices. However, there is a dearth of prior studies on examining the factors that influence the cloud computing adoption among Australian SMEs. To fill the empirical vacuum, this research-in-progress proposes an integrated framework for examining the determinants of cloud computing service adoption with the consideration of the unique characteristics of Australian SMEs, such as relatively low adoption of cloud computing services, less innovative, and limited knowledge about cloud computing and its benefits and hindrances. To this end, we are conducting consecutive studies to investigate this research issue. An exploratory interview study will be applied to observe and verify the characteristics of Australian SMEs toward the cloud computing adoption. This is followed by an organisational level survey that examines the effects of determinants on cloud computing adoption. Finally, a decision model for cloud computing adoption among Australian SMEs will be developed by using a Multi Criteria Decision Approach (MCDA) through rating, prioritising, and ranking of various criteria and alternatives available to the decision makers. Adopting the mixed-method research fashion, this research-in-progress intends to make significant implications to scholars and practitioners alike in the cloud computing research and applications areas.

**Keywords**

ICT Adoption, Small and Medium Enterprises (SMEs), Cloud Computing, Theories [Diffusion of Innovation (DOI), Technology-Orgization-Environment Framework (TOE), Agent Network Theory (ANT)].

## 1. INTRODUCTION

Cloud computing is a new emerging technology for the future ICT (Information and Communication Technology). Over the years there has been increasing interest in its use (El-Gazzar 2014) It is changing the computing perspective. The cloud computing has been ranked as the top 5 of most the influential initiatives for organizations in developed countries such as the United States of America and Australia (Baty and Stone 2011; CIO 2013). It has also been rec-ommended to organizations in developing countries such as African countries in order to enable them to follow the new trend in technologies through affordable technological solutions (Greengard 2010).

From observation and according to experts in this field (Carcary et al. 2014; Dillon and Vossen 2014), SMEs (Small and Medium-sized Enterprises) sector is one of the economical entities that can benefit highly from the use of cloud computing services. Cloud computing can support elastic and efficient business models (Benton 2010). This suggests that SMEs can grasp these kinds of features and also other kind of services such as storage infrastructure and computational capabilities offered by cloud computing to increase its business performance and productivity. In practice, however, reports show that there has been a slow adoption of cloud computing services among SMEs (Khajeh-Hosseini et al. 2010). Security concerns are one of the main hindrances to the adoption of this technology (Kim et al. 2009). Security is not only a concern for large organisations but for all organisation types and sizes, including SMEs (Kim et al. 2009). SMEs have much sensitive data that needs to be protected, such as quotations to their customers, financial details, company databases, trade secrets, email accounts, research findings, confidential research, feasibility studies etc. (Misra and Mondal 2011). In addition to security issue, Koehler et al. (2010) found that reliability is also one of the main obstacles to the adoption of cloud computing. Furthermore, a study conducted by (Catteddu and Hogben 2009) found that the main obstacles to the adoption of cloud computing are unwillingness due to the large capital investment, privacy, security concerns, availability and integrity issues, and confidentiality of information. As a result, there are several barriers for the adoption of cloud computing among SMEs (Misra and Mondal 2011).

Although prior studies have been done to examine the direct influence of technological innovation attributes or other contextual elements of cloud computing adoption among SMEs, few investigations focus on the adoption of cloud computing among SMEs from a holistic view of technology, organization and environment perspectives (Martins et al. 2014). The technological aspects of cloud computing are not the only thing to consider. It is necessary to understand the whole system from the provision of services to deployment (Creeger 2009). In this regard, this research-in-progress tries to address the following research questions:

1. What are the determinants that influence the decision of an SME to adopt cloud computing technology?
2. If the decision is to adopt the cloud computing technology, what factors determine the extent of such adoption?

To this end, this research-in-progress investigates the socio-technical factors that influence SMEs in their decision about the adoption of cloud computing services in Australia. Based on theory of diffusion of innovation, technology-orgisation-environment framework and agent network theory, we propose a multi-perspective framework that originated from three relevant grounded theories to investigate the determinants of cloud computing adoption among SMEs in Australia. In addition, we apply a mixed-method research approach to empirically validate these determinants. An exploratory interview study will be applied to observe/verify the characteristics of Australian SMEs toward the cloud computing adoption. Next, an organisational level survey study will be conducted to examine the effects of determinants on cloud computing adoption. Finally, a refined decision model for cloud computing adoption among Australian SMEs will be obtained from a multiple criteria ranking and prioritizing decision study. In this regard, this research-in-progress intends to make significant implications to scholars and practitioners alike in the cloud computing research area.

The reminder of this research-in-progress is organized as follow: we review prior literature on cloud computing and Australian SMEs in section 2. In section 3, we propose and discuss the theoretical foundation of this research-in-progress. This is followed by the research model and hypotheses development in section 4. After that, we propose the consecutive studies for examining the research question of this paper. The expected contributions and implications are discussed at the end of this research-in-progress.

## 2. RELATED WORK

### 2.1 Cloud Computing Adoption

Cloud computing, an emerging concept, has received much attention in both the economic and academic fields. There are several additional definitions of cloud computing in the literature from both practical and academic perspectives. For this paper we use the definition that was introduced by the US National Institute of Standards and Technology (NIST) as:

*"... A model for enabling ubiquitous, convenient, on-demand network access to a shared pool of configurable computing resources (e.g. networks, servers, storage, applications, and services) that can be rapidly provisioned and released with minimal management effort or service provider interaction" (Mell and Grance 2011).*

For the purpose of this research, a definition from The National Institute of Standards and Technology (NIST) will be our referencing basis as it is found to be more detailed in describing the framework and satisfying the stipulated objectives that have gained universal acceptance across business, industry, and research. Basically, cloud computing service models have five common characteristics that make it unique in comparison with other computing resources: (1) Providing On-demand self-service such as storage; (2) Broad network access to heterogeneous platforms such as mobiles phones and computers; (3) Wide range of resource pooling multi-tenant concept via physical or virtual mechanisms on consumer demand; (4) Inward and outward scalability of resources based on consumer demand; (5) Measured services for efficiency and optimization (Mell and Grance 2011).

There have been limited investigations conducted to address the adoption of cloud computing from the organisational side (El-Gazzar 2014) Table 1: presenting some seminal prior studies. Researches have primarily addressed the direct influence of technological innovation attributes or other contextual elements (Martins et al. 2014). It is believed that methodologically, theoretically, and contextually this area requires further investigation.

| Dependent variables | Independent variables researched | Theoretical perspective | Context/Unit level | Sources |
|---|---|---|---|---|
| Cloud computing | Perceived benefits, business concerns, IT capability, external pressure, and firm size. | TOE | 200 Taiwanese companies/ Organisational | (Hsu et al. 2014) |
| | Barriers and benefits | - | Survey of 94 SMEs in Spain/ Organisational | (Trigueros-Preciado et al. 2013) |
| | Technology (relative advantage, Tech complexity, Tech compatibility), organisation (top management support, firm size, IT expertise of business users) and environment (competition intensity, regulatory environment) | TOE | A survey study of 24 global organisations in different industries/ Organisational | (Borgman et al. 2013) |
| | Business process com- | Innovation Dif- | Survey on Manu- | (Wu et al. |

| | plexity, entrepreneurial culture, compatibility, application functionality. | fusion Theory & Information Processing View | facturing and retail firms/ Organisational | 2013) |
|---|---|---|---|---|
| | Perceived technology barriers, perceived environment barriers, perceived benefits. | TOE | Used secondary sourced data/ Organisational | (Nkhoma et al. 2013) |

*Table 1 Cloud computing adoption, investigated variables, and research context.*

## 2.2 Australian SMEs

According to the Australian Bureau of Statistics (ABS) definition, there are three types of SMEs: (1). Micro businesses being those with 0-4 employees; (2). Small businesses with 0-19 employees; and (3). Medium businesses are those with 20-199 employees (ABS 2001). This definition fits into our study context and objective.

SMEs play a vigorous role in the Australian economy, they account for 99.75 in business economy and employ over 70 percent of the country workforce (ABS 2013). The sector contribute to more than AU$ 480 billion to the country economy (Clark et al. 2011). More than 80 percent of businesses in Sydney are small firms, and contribute more than AU$ 25 billion/year which makes around 25 percent of the city's economic production.

There is a remarkable gap in the adoption of technology between large corporations and SME (Pick and Azari 2008). The SME sector contributes significantly to the Australian economy and it is aligned with the country strategic objectives (ABS 2013). Innovation in products, services, and processes is the key for lifting Australian; this can be achieved through adoption of new technologies. Australia has limited contribution to the global innovations (Daley et al. 2013). Australia is behind other counties in the quality and prices of International and local network connections. For example, it is behind all Asian developed countries in cloud readiness and rated low on global Internet connectivity (Asia-Cloud-Computing-Association 2012). Australia is also behind other OECD (Organisation for Economic Co-operation and Development) countries on domestic broadband speed and prices (OECD 2013b).

Australian ICT adoption is higher than many of the OECD peer countries (OECD 2013b), however adoption is less among SMEs than among large organisations (ABS 2013). Many Australian SMEs do not have enough knowledge of what the term 'cloud computing' means and are not aware of its benefits MYOB (2012). In an earlier survey in 2011, Optus found that 59 percent of SMEs are not aware or sure of cloud computing (Optus 2011). The Australian computer society stated that the ACMA survey showed 52 percent of respondents have concerns about privacy and lack of confidentiality as cloud computing is more exposed to privacy and security breaches than other computing paradigms (ACMA 2014). Security and privacy of cloud computing are major concerns of Australian SMEs (Senarathna et al. 2014). Network quality is an important factor in fostering the adoption of cloud services. Slow and unreliable connections are also a problem for cloud services (Australian-Government 2012). The Australian authorities are responsible for ensuring the availability of reliable fixed and mobile network connections, low-latency uploads and downloads, and adequate international communication to support the use of cloud services by SMEs. The obstacles associated with regional network coverage and other hindrances also need to be overcome. The government recognizes all these issues (Australian-Government 2012), but positive action must be taken for the benefit of the businesses and the economy.

Cloud computing offers opportunities to Australian SMEs to acquire advanced and flexible IT services at a relatively reasonable cost. These services that require low investment can have the potential to leverage the competitiveness of the sector and increases its productivity and efficiency. The industry is very big and the benefits and challenges of cloud computing are realised. Therefore, we believe that understanding the influential factors that impact the adoption of cloud computing is important in order to provide the concerned stakeholders in this study with more informative decision.

## 3. CONCEPTUAL FRAMEWORK

The majority of the research into the adoption and diffusion of ICT studies the two levels: the organisational level and the user level (Choudrie and Dwivedi 2005). Choudrie and Dwivedi (2005) indicated the importance of considering the context, adoption stages, and the usefulness of a selected method in their research formulation. Because SMEs were the context of their research, the organisational level was their parameter of analysis. This means that the theories associated with the investigation of individual level choices were not considered. Agent network theory (ANT) was used as the theoretical framework for analysing the qualitative data collected through the literature, interviews, and other techniques used in ethnography (Deering et al. 2012). Hence, in this research, the qualitative study was one of the methods used in combination with quantitative study. This is the 1$^{st}$ phase of the research and it followed exploratory research design. Interviews were conducted to obtain an in-depth understanding of the issues related to the cloud computing adoption. The outcome of this study will be the feed of a larger scale survey study.

From the theoretical consideration point of view, it is important to understand that there are three distinguishable dimensions in investigation technological adoption: users level dimensions, firm level, and market/innovation level. The most influential theories that have been used to study these levels are TAM (Technology Acceptance Model), DOI (Diffusion of Innovation), TRA (Theory of Reasoned Action), TPB (Theory of Planned Behaviour), UTAUT (Unified Theory of Acceptance and Use of Technology). For example, TAM was widely used to study ICT innovation phenomena at individual levels (RUI 2007), whereas, DOI is mainly used to study market levels technological innovation but it has limitations in considering the environmental perspectives in organisational adoption of a new innovation, mainly because of its technical perspective (RUI 2007). Technology-Orgization-Environment Framework (TOE) consists of three main contexts- Technological context, Organisational context, and Environmental Context. It is broader, contain organisational elements, and fit to our research (Tornatzky et al. 1990). By combining DOI to TOE we reduce some of the limitations of DOI and present a more comprehensive research framework. In addition, the third theory, ANT, with its association in explaining the relationships among people, objects and organisations is believed to be appropriate to the socio-technical environments of cloud computing adoptions where both aspects "technical" and "social" have importance understanding of the adoption process.

TOE and ANT provided researchers in the field of IS (information Systems) innovation and various other fields with a theoretical foundation for their studies. In this research, an integration of the three theories: DOI, TOE, and ANT have been developed to provide a framework for the investigation of the adoption of cloud computing. This research model addresses the limitations previous studies of the non-comprehensive framework for ICT innovation adoption. Most of the previous academic and industrial studies including the presented studies in Table1 considered technological adoption factors, some of them considered organisational factors, some used secondary data, other studies considered wider dimensions but targeted a smaller scale of population, and noticeably these studies were limited in using comprehensive methodological research approaches. We believe that our proposed method will be able to obtain more conclusive results.

## 4. RESEARCH MODEL AND HYPOTHESES

The proposed integrative research framework is shown in Fig. 1. As discussed, this presents a more holistic model for understanding the diffusion of innovation topics and, as such, it will be the foundation of this study.

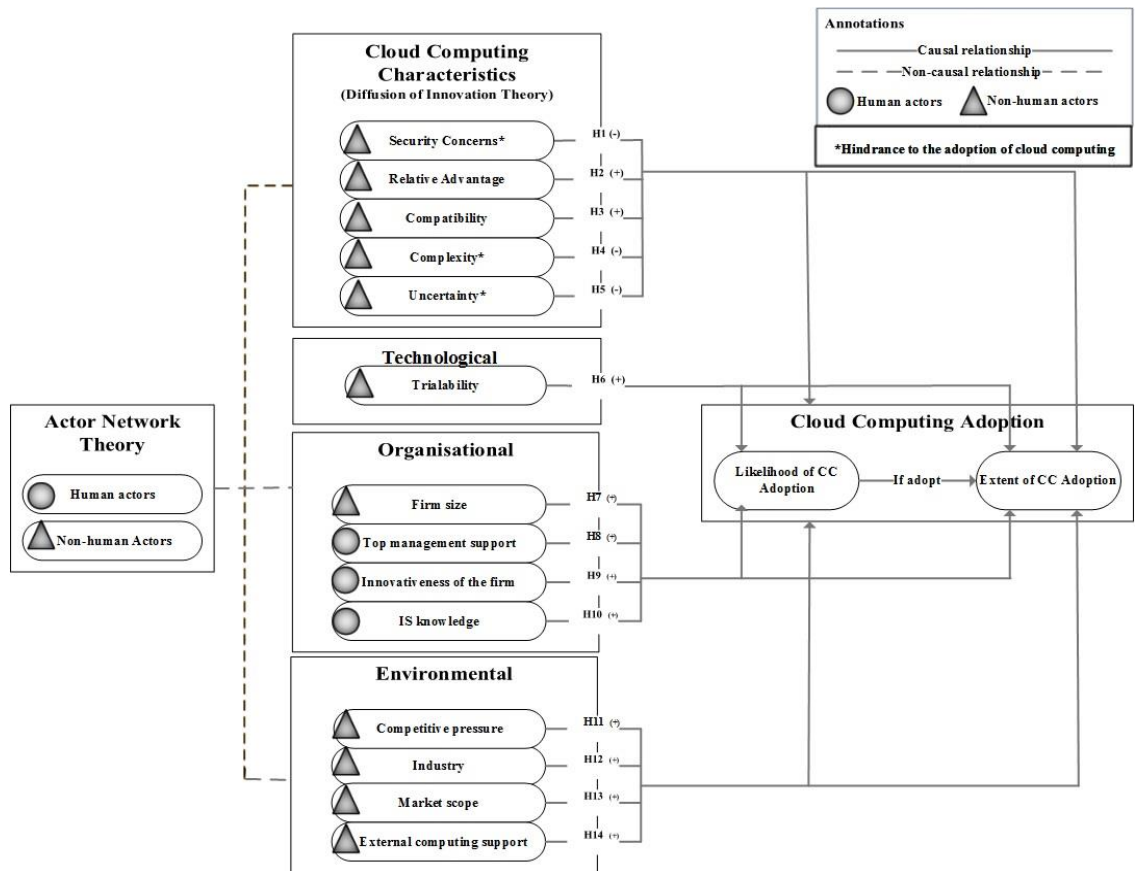

*Figure 1 An integrated model for adoption of cloud computing by SMEs*

The fourteen relevant factors are classified into five groups. The five groups are (1) technological factors, (2) organisational factors, (3) environmental factors that impact the adoption of cloud computing among SMEs. The other two outer groups which are (4) human and (5) non-human actors do not have a causal relationship in other factors. However, they are used to help in better understanding the research situation and appreciate the network of the actor's elements whether human or non-human to the overall phenomena. As it has been indicated briefly earlier, the constructs have been mainly derived from three theories: DOI, TOE, and ANT. The factors have been modified based on the prior studies (Borgman et al. 2013; Hsu et al. 2014; Nkhoma et al. 2013; Trigueros-Preciado et al. 2013; Wu et al. 2013). Careful consideration has been given in making these constructs compatible with the objective of this study and the contextual characteristics of SMEs. These considerations involved selecting the most influential constructs that have been reported in similar studies with closer context like other OECD countries and the studies that have been conducted in Asian developed countries like China and Singapore. With a limited literature in cloud computing adoption in SMEs, ICT innovation adoption knowledge was the foundation of the study and particularly Thong (1999) study was one of the main reference of this study.

Technological factors have been derived mainly from Rogers' DOI theory. The theory was introduced in 1962, Roger listed four attributes of innovation namely: relative advantage, complexity, compatibility, and trialability (Rogers 2003). A further study conducted by Tornatzky and Klein (1982) acknowledged that relative advantage, complexity and compatibility are elements of significant impact on innovation. As we know, major cloud computing providers such as Amazon and Microsoft offers trial versions of their cloud services to its clients. Therefore, the trialability attribute is essential and it is included as an influential construct in our study. On the other hands, organisational factors are primarily associated with the characteristics of the organisation itself that impact on the adoption decision. These include: company's information resources and its employees' knowledge. Environmental factor are the external aspects that have influence on organisation's innovation decisions. Human factors are the responsible individuals in making innovation decisions. (Thong and Yap 1995) revealed that SMEs, decision makers have remarkable influence in making innovation decision. Non-human factors are all other influential factors that are not related to human beings. More details about all the factors will be presented in the hypotheses development section.

Several factors in the literature are believed to have a significant impact on cloud computing adoption by Australian SMEs. Although numerous other factors exist, the constructs at this stage are mainly derived from widely acknowledged theories and from some of the presented academic and industrial literature from Australia and from other settings such Asian developed countries and other OECD which are believed to be relevant and to have impact on Australian SMEs. The model applied a multidimensional approach and presented a framework for investigating the research topic.

## 4.1 Hypotheses of the Technological Factors

- **Security concerns**

A security breach is the occurrence of breach in which a private or public organisation loses data, individual records, or other critical information (Bishop 2012). The shared multi-tenant environment, which is offered by the cloud computing paradigm increases security concerns (Schneiderman 2011).

*H1: Perceived security and privacy concerns of cloud computing will be negatively related to the likelihood of cloud computing adoption.*

- **Relative advantage**

The relative advantage as defined by Rogers (2003) is "the degree to which an innovation is perceived as being better than the idea it supersedes". In this study, the innovation referred to is cloud computing and the superseded idea is the other computing paradigms.

*H2: Perceived relative advantage of cloud computing will be positively related to the likelihood of cloud computing adoption.*

- **Compatibility**

Compatibility is defined as "The degree to which an innovation is perceived as consistent with the existing values, past experiences, and needs of potential adopters" (Rogers 2003). In this study, compatibility will be investigated mainly in the business operations as well as the compatibility with its technological resources.

*H3: Perceived compatibility of cloud computing will be positively related to the likelihood of cloud computing adoption.*

- **Complexity**

This attribute is adopted from Roger's DOI theory and it is defined as "The degree to which an innovation is perceived as relatively difficult to understand and use" (Rogers 2003). It is believed the same can be relevant to cloud computing.

*H4: Perceived complexity of cloud computing will be negatively related to the likelihood of cloud computing adoption.*

- **Uncertainty**

Uncertainty is defined as the extent to which the consequences of using an innovation are insecure or uncertain (Fuchs 2005). It could be a significant barrier for SMEs to adopt cloud computing.

*H5*: *The perceived uncertainty of cloud computing will be negatively related to the likelihood of cloud computing adoption.*

- **Trialability**

Roger defined it as "The degree to which an innovation may be experimented with on a limited basis" (Rogers 2003). This probably can have a significant contribution in the adoption of cloud computing in conjunction with other attributes that are believed to be crucial to be investigated such as relative advantages, complexity, and compatibility.

*H6*: *Trialling cloud services before adoption will be positively related to the likelihood of cloud computing adoption.*

## 4.2 Hypotheses of the Organisational Factors

- **Firm size**

Due to economic factors, the size of the firm is one of the major factors of IT adoption (Pan and Jang 2008). This means that it is an essential factor for cloud computing adoption.

*H7*: *Firm size will be positively related to the likelihood of cloud computing adoption.*

- **Top management support**

Top management support is an essential criterion for providing sufficient resources in the adoption of new innovation and to provide necessary support required for the innovation, re-engineering, and change process (Wang et al. 2010).

*H8*: *Top management support will be positively related to the likelihood of cloud computing adoption.*

- **Innovativeness of the firm**

Innovativeness is the degree of willingness to take risk and try new solutions or technologies that have not been tried or tested before (Thong and Yap 1995). This is relevant to cloud computing as a relatively new innovation

*H9*: *Innovativeness of the firm will be positively related to the likelihood of cloud computing adoption.*

- **IS knowledge**

In general, small businesses are lacking in specialised IS knowledge and technical skills (Plomp et al. 2014). Thong (1999) found that CEO's technological knowledge have positive influence in the diffusion of information systems. This study will investigate if the decision makers characteristics in innovation and knowledge will have impact in adoption of cloud computing.

*H10*: *IS knowledge will be positively related to the likelihood of cloud computing adoption.*

### 4.3 Hypotheses of the Environmental Factors

- **Competitive pressure**

The term refers to the extent of competition intensity among firms in a single industry within which an organisation functioning (Thong & Yap, 1995). Competition increases the uncertainties in the marketplace and therefore originate an increase in the rate of innovation adoption (Ettlie 1983). Therefore, for small businesses this can become a mandatory and strategic move toward new technologies to gain competitive grounds over rivals.

*H11*: *Competitive pressure will be positively related to the likelihood of cloud computing adoption.*

- **Industry**

Industry type can have influence on the firm's decision about the adoption of new technologies (Jeyaraj et al. 2006). Different industry sectors adopt cloud computing services at different rate (Low et al. 2011).

*H12*: *The industry within which SMEs operate will be positively related to the likelihood of cloud computing adoption.*

- **Market scope**

The larger the size of the market of an individual business, the greater the need for more information and communication technologies in order to remain competitive in such large markets (Hitt 1999).

*H13*: *SMEs with wider market scope will be positively related to the likelihood of cloud computing adoption.*

- **External computing support**

This attribute is defined as "the availability of support for implementing and using an information system" (Premkumar and Roberts 1999). It discusses how the external support such as training, customer service, and technical support that are offered by cloud providers can influence in increasing the cloud services adoption by SMEs.

*H14*: *External computing support will be positively related to the likelihood of cloud computing adoption.*

## 5. METHODOLOGY

### 5.1 Study Design

This research contains three consecutive studies. The first study is an exploratory interview study. It will be applied to observe and verify the characteristics of Australian SMEs toward the cloud computing adoption. This will be followed by an organisational level survey study, which will examine the effects of determinants on cloud computing adoption. In the third study, we will design a decision model for cloud computing adoption through ranking and prioritizing of criteria and alternatives obtained from the previous qualitative and quantitative studies.

### 5.1.1 Interview study

This is 1st phase of the study. It will take an exploratory orientation. It is targeting to interview the decision makers of 15 organisations (4 cloud computing services providers, 6 cloud computing adopters, 2 prospectors, 3 companies not intend to adopt cloud computing). The method used and is semi-structured interviews. The purpose of this study is to investigate the influential factors that affect the adoption decision of cloud computing services empirically and then to feed them in a larger scale quantitative study.

### 5.1.2 Organisational level survey

This is the 2nd phase of the study and it is depending on the inputs from the 1st phase. A directory list of organisations and SMEs from different sectors (i.e. manufacturing, retailing, service, etc.) will be obtained from "Orbis" (a global comprehensive database of companies which will contain the required information). A formula presented by Tabachnick and Fidell (2007) for calculating sample size requirements will be implemented, taking into consideration the number of independent variables to be used: $N > 50 + 8m$ (where $m$ = number of independent variables). Therefore, depending on the number of interdependent variables, the number of required cases will be calculated. The firm size considered taken into account is: 0-199 employees; region: Australia; a simple random sample for selection of firms will be used. This study will use mixed method research for collecting data. According to (Saunders et al. 2011), this will provide good confidence in the final results.

### 5.1.3 Designing decision model

This is the last stage of the research and it will use the data obtained from both the previous two studies to decision the decision model. In addition, the study will involve 4 firms in the decision process as well. One of these firms will be a cloud computing services provider. The other 3 firms will be either users or firms that intend to use these services. Designing methodology requires having two main components: criteria and alternatives. These components need to be modified to fit into the model design and structure. The model will be designed using a Multi Criteria Decision Approach (MCDA). This approach will produce a model that can perform (1) rating (2) ranking (3) prioritizing (4) selection of (a) cloud computing services (b) deployment models (c) services providers

### 5.2 Statistical Techniques and Research Methods

The statistical techniques that will be used in this are factor analysis, multivariate analysis, frequencies, correlations and multiple regressions. Table 2 summarises the main research methodology for this study.

| **Research Philosophy** | **Research Approach** | **Research Strategy** | **Research Method** |
|---|---|---|---|
| Post-positivism | Quantitative | Experiment | Documents |
| Positivism | Qualitative | Survey | Interviews |
| Realism | Mixed Methods | Case Study | Direct Observations |
| Constructivism | | Historical | Field Notes |
| | | | Archive |

*Table 2 Research methodology selection.*

### 5.3 Study Measurement

This study will measure the two criteria of validity and reliability. These measures are usually difficult to achieve in qualitative research (Denzin and Lincoln 2009) but, the measurement will help in provide a checking tool for enhancing the quality of the research. Pre-tests and pilot studies will be used to validate the applicability and validity of the survey academically and practically. Principal component factor analysis (PCA) tool will be used for all items in the survey questionnaire. This is a useful tool in assessing the convergent and discriminant validity (Pallant 2013). This study will follow the recommendation of (Hinton 2004) has for the four cut-off points for reliability:- excellent reliability (0.90 and above), high reliability (0.70-0.90), moderate reliability (0.50- 0.70) and low reliability (0.50 and below).

## 6. PRELIMINARY FINDINGS

At this stage seven organisations have been interviewed using semi-structured approach including three cloud computing services providers, two SMEs cloud services adopters and two SMEs not adopting cloud computing. This first run of the qualitative study found that all the proposed factors in the research model are supported by the firms except for the 'competitive pressure' as it was not been considered as a significant factor in the cloud computing adoption decision in comparison with other more influential factors.

Security found to be a concern with the majority of the participants with their different adoption stages. This found to be mainly depending on the industry type and the sensitivity of data. In addition, two new factors, cost savings or service costs and privacy due to geo-restriction were identified and found to be important for SMEs when deciding in implementing cloud computing services. Costs of services have high impact on the decision of SMEs migration to cloud due to their financial constraints. It is easier for SMEs to implement cloud computing services if they find that a total cost of ownership for cloud services provide them with more and better advanced services than having in-house computing resources. Privacy due to geo-restriction was found to be a crucial factor and this is due to firm's preferences for their data to be stored locally in Australian boundaries as they trust the regulation of their own country and are not confident with other countries jurisdiction due to the absence of a global governance of cloud computing.

## 7. EXPECTED CONTRIBUTIONS

The study is expected to deliver the following two contributions:

### 7.1 Theoretical Contribution:

- This study will have insights for research. To the best of our knowledge, this is one of the first rigorous attempts that examined cloud computing adoption in Australian SMEs from a theoretical and empirical perspective. The model was developed based on an integration of various perspectives using the ICT innovation literature as a basis. The model will be then empirically tested using multivariate statistical techniques for analysis. Bivariate correlation analysis on SMEs is common in prior research. Multivariate analysis consider for the interdependencies between constructs that are not accounted for with bivariate analysis.

- This research extends the usefulness of the TOE framework, DOI theory, and ANT theory in providing guidelines on organisation intention to adopt cloud computing services. It will also provide a validated measurement model for the adoption of cloud computing.

### 7.2 Practical Contributions:

- Implications for technology consultants and cloud services providers including software vendors: understanding the influential factors helps in design strategies and provides better services and products.
- Implications for organisation's decision maker: Help decision makers to evaluate possible adoption and increase their awareness about factors that influence results by provide them with a set of valid and reliable measurements for evaluating the determinants that influence adoption.
- Implications for government and policy makers: providing an understanding of the factors that influence the adoption of cloud computing which will allow government to take various measures to refine and enhance business processes and to increase connectivity among SMEs through implementing proposer legislations on privacy and data protection or enhancing the existing legislations for improving the performance of the SMEs sector.

Despite the limitations that might occur, this study will develop and test an integrated model of cloud computing adoption in SMEs based on a review of the ICT innovation literature and empirical studies and present some major determinants of cloud computing adoption in SMEs.

## Acknowledgement

The authors gratefully acknowledge the help of Dr Madeleine Strong Cincotta in the final language editing of this paper.